\documentclass[useAMS,usenatbib]{mn2e}
\usepackage{amsmath,amssymb}
\usepackage{graphicx}

\title[stochastic background of gravitational waves]{Stochastic background of gravitational waves generated by pre-galactic black holes}
\author[E. S. Pereira and O. D. Miranda]{Eduardo S. Pereira\thanks{E-mail:
duducosmo@das.inpe.br} and Oswaldo D. Miranda\thanks{E-mail: oswaldo@das.inpe.br}\\
INPE - Instituto Nacional de Pesquisas Espaciais - Divis\~{a}o de Astrof\'{i}sica,\\
Av. dos Astronautas 1758, S\~{a}o Jos\'{e} dos Campos, 12227-010 SP, Brazil}
\begin{document}

\date{Accepted  . Received ; in original form }

\pagerange{\pageref{firstpage}--\pageref{lastpage}} \pubyear{2009}

\maketitle

\label{firstpage}

\begin{abstract}

In this work, we consider the stochastic background of gravitational waves (SBGWs) produced by pre-galactic stars, which form black holes in scenarios of structure formation. The calculation is performed in the framework of hierarchical structure formation using a Press-Schechter-like formalism. Our model reproduces the observed star formation rate at redshifts $z\lesssim 6.5$. The signal predicted in this work is below the sensitivity of the first generation of detectors but could be detectable by the next generation of ground-based interferometers. Specifically, correlating two coincident advanced LIGO detectors (LIGO III interferometers), the expected signal-to-noise-ratio (S/N) could be as high as 90 (10) for stars forming at redshift $z\simeq 20$ with a Salpeter initial mass function with slope $x=0.35$ ($1.35$), and if the efficiency of generation of gravitational waves, namely, $\epsilon_{\rm GW}$ is close to the maximum value $\sim 7\times 10^{-4}$. However, the sensitivity of the future third generation of detectors as, for example, the European antenna EGO could be high enough to produce ${\rm (S/N)}>3$ same with $\epsilon_{\rm GW}\sim 2\times 10^{-5}$. We also discuss what astrophysical information could be derived from a positive (or even negative) detection of the SBGWs investigated here.

\end{abstract}


\section[]{Introduction}

Gravitational waves (GWs) are a natural consequence of Einstein's theory of general relativity (GR). GWs will open a new astronomical window for the study of the Universe transforming the research in GR into an observational/theoretical study. In particular, the opening of the full electromagnetic spectrum to astronomical observation during the last century expanded our comprehension of the Universe. In this century, observations across the gravitational wave spectrum will provide a wealth of new knowledge, including the possibility of studying the period when the first stars were formed in the Universe in the end of the so-called `dark ages'.

The information provided by GWs is different when compared to that provided by electromagnetic waves. GWs carry detailed information on the coherent bulk motions of matter, such as those produced by the collapse of stellar cores generating, for example, black hole remnants. On the other hand, electromagnetic waves are usually an incoherent superposition of emissions from individual atoms, molecules, and charged particles.

Because of the fact that GWs are produced by a large variety of astrophysical sources and cosmological phenomena, it is quite probable that the Universe is pervaded by a background of such waves. Collapse of Population II and III stars, phase transitions in the early Universe, cosmic strings, and a variety of binary stars are examples of sources that could produce such a putative background of GWs (see, e.g., \citealt{mag1,d4,d5,sand1,suwa,gio} among others).

Note that the indirect evidence for the existence of gravitational waves came first from observations of the orbital decay of the Hulse-Taylor binary pulsar \citep{hulse1,hulse2,hulse3}. Direct detection though and analysis of gravitational-wave sources are expected to provide a unique insight to one of the least understood of the fundamental forces   \citep{belcz}. They will also allow us to investigate the physical properties of objects that do not emit any electromagnetic radiation as for example isolated black holes.

A number of interferometers designed for gravitational wave detection are currently in operation, being developed, or planned. In particular, the high frequency part of the gravitational wave spectrum ($10{\rm Hz} \lesssim f \lesssim 10^4{\rm Hz}$) is open today through the pioneering efforts of the first-generation ground-based interferometers such as LIGO. While detections from this first generation of detectors are likely to be rare, the advanced LIGO upgrade may detect, among others, the stochastic signal generated by a population of pre-galactic stars.

Thus, in the future, it may be possible to use GWs as a tool for studying the star formation at high redshifts. In particular, from the theoretical point of view, it can be found in the literature several works discussing this possibility. For example, the gravitational wave background (GWB) generated from the core collapse supernovae resulting in black holes at high redshifts has been discussed by \citet{f2,d3,d5} among others. On the other hand, the calculation made specifically for Population III supernovae  resulting in black holes is presented in \citet{d4}.

More recently, \citet{sand1} calculated the GWB from Population III stars with the cosmic star formation history in the framework of hierarchical structure formation. On the other hand, \citet{suwa} presented the GWB spectrum of Population III stars by calculating the GW waveforms based on results of hydrodynamic core-collapsed simulations (see also \citealt{suwa2}). It is worth stressing that in all of these works, one of the most important parameters responsible to characterize the GWB is the cosmic star formation rate (CSFR). 

Concerning to the CSFR at high redshift, our knowledge is mainly based on numerical simulations performed by hydrodynamical codes in a $\Lambda$-CDM cosmology. Certainly, these simulations must reproduce the observable Universe at redshifts $z \lesssim 6$. In particular, the evidence for the existence of a large star formation at high redshift comes from, among others, the Gunn-Peterson effect \citep{gunn} and from the metallicity of $\sim 10^{-2} Z_{\odot}$ found in ${\rm high}-z$ Ly$\alpha$ forest clouds \citep{songaila,ellison}.

These results are consistent with a stellar population formed at $z \gtrsim 5$ \citep{venk}. However, measuring the CSFR from observations requires a number of assumptions, with the form of the dust obscuration corrections and the stellar initial mass function \citep{kroupa,wilkins}.

Our main goal in the present paper is to discuss how the detection of a GWB could be used to give us some insight on the CSFR. This kind of study could also be used to constrain the fraction of massive stars that generates black holes at high redshift, and the efficiency of production of GWs by black holes whose distribution function is presently unknown. To do so, we use a hierarchical structure formation model similar to that developed by \citet{d1}.

However, in our model the CSFR is obtained in a self-consistent way. That means, we solve the equation governing the total gas density taking into account the baryon accretion rate, treated as a infall term, and the lifetime of the stars formed in the dark halos.

The paper is organized as follows. In Section 2, we present the Press-Schechter-like (PS) formalism used to determine the comoving abundance of collapsed dark matter halos. In Section 3, we discuss how to obtain the CSFR from the hierarchical model. In Section 4, we present the formalism used to characterize the GWB. Section 5 presents our conclusions.

\section[]{Hierarchical formation scenario}

Press and Schechter (hereafter PS) heuristically derived a mass function for bound virialized objects in 1974 \citep{p2}. The basic idea of the PS approach is define halos as concentrations of mass that have already left the linear regime by crossing the threshold $\delta_{\rm c}$ for non-linear collapse. Given a power spectrum and a window function, it should then be relatively straightforward to calculate the halo mass function as a function of the mass and redshift. However, it is worth stressing that the exact definition of the mass function, e.g., integrated versus differential form or count versus number density, varies widely in the literature. To characterize different fits, it can be introduced the scale differential mass function $f(\sigma,z)$ \citep{j1} defined as a fraction of the total mass per $\ln \sigma^{-1}$ that belongs to halos. That is,

\begin{equation}
f(\sigma,z)\equiv\frac{d\rho/\rho_{\rm B}}{d\ln\sigma^{-1}}=\frac{M}{\rho_{\rm B}(z)}\frac{dn(M,z)}{d\ln[\sigma^{-1}(M,z)]}.
\end{equation}

\noindent Where $n(M,z)$ is the number density of halos with mass $M$, $\rho_{\rm B}(z)$ is the background density at redshift $z$, and $\sigma(M,z)$ is the variance of the linear density field. As pointed out by \citet{j1}, this definition of the mass function has the advantage that it does not explicitly depend on redshift, power spectrum, or cosmology; all of these are contained in $\sigma(M,z)$ (see also \citealt{luk}).

To calculate $\sigma(M,z)$, the power spectrum $P(k)$ is smoothed with a spherical top-hat filter function of radius $R$, which on average encloses a mass $M$ $(R=[3M/4\pi\rho_{\rm B}(z)]^{1/3})$. Thus,

\begin{equation}
\sigma^{2}(M,z) = \frac{D^{2}(z)}{2\pi^{2}} \int_{0}^{\infty}{k^{2}P(k)W^{2}(k,M)dk},
\end{equation}

\noindent where $W(k,M)$ is the top-hat filter:

\begin{equation}
W(k,M) = \frac{3}{(kR)^{3}}[\sin(kR)-k R\cos(kR)],
\end{equation}

\noindent and the redshift dependence enters only through the growth factor $D(z)$. 

Then,

\begin{equation}
\sigma(M,z)=\sigma(M,0)D(z).
\end{equation}

In the more general case of a Universe with matter and a cosmological constant, the exact solution for the growth function is well approximated by \citep{c2}:

\begin{equation}
D(a)\approx \frac{5 \Omega_{\rm m}(a)\ a}{2[1- \Omega_{\Lambda}(a)+\Omega_{\rm m}^{4/7}+\frac{1}{2}\Omega_{\rm m}(a)]},
\end{equation}

\noindent where the relative density of the ${\rm i}-$component is given by $\Omega_{\rm i}=\rho_{\rm i}/\rho_{\rm c}$, and `${\rm i}$' applying for baryons (b), dark energy ($\Lambda)$, and total matter (m), while $a=1/(1+z)$ is the cosmological scale factor.

As usual, the primordial power spectrum is assumed to have a power law dependence on scale, that is, $P(k)\propto k^{n}$. For a scale-invariant spectrum the spectral index is $n=1$. The rate at which fluctuations grow on different scales is determined by an interplay between self-gravitation, pressure support and damping processes. These effects lead to a modification of the form of the primordial power spectrum that is expressed in terms of a transfer function $T(k)$. Thus, we have:

\begin{equation}
P(k) = BkT(k),
\end{equation}
\noindent where the normalization factor $B$ is determined observationally.

For the transfer function, we consider \citep{e1} :

\begin{equation}
T(k) = \frac{1}{\{1+[ak+(bk)^{3/2}+(ck)^{2}]^{\nu}\}^{2/\nu}},
\end{equation}

\noindent with $\nu = 1.13$, $a = (6.4 /\Gamma) h^{-1}\rm{Mpc}$, $b= (3.0/ \Gamma)h^{-1}\rm{Mpc}$, $c = (1.7 /\Gamma) h^{-1}\rm{Mpc}$, and $\Gamma = \Omega_{\rm m} h\,\,{\rm e}^{-\Omega_{\rm b}(1+\sqrt{2h}/\Omega_{\rm m})}$ is the so-called shape parameter of the power spectrum \citep{b2,p1}.

We use throughout this work the mass function fit proposed by \citet{st}. That is,

\begin{equation}
f_{\rm ST}(\sigma) = 0.3222 \sqrt{\frac{2a}{\pi}} \frac{\delta_{\rm c}}{\sigma} \exp{\left(-\frac{a \delta_{c}^{2}}{2 \sigma^{2}} \right)} \left[1+\left(\frac{\sigma^{2}}{a\delta_{\rm c}^{2}}\right)^{p}\right]\label{est},
\end{equation}

\noindent where $a = 0.707$ and $p=0.3$. 

At redshift $z$, the comoving density of dark matter halos in the mass range $[M,M+dM]$ is $f_{\rm ST}(\sigma)dM$, with (see, in particular, \citealt{d1})

\begin{equation}
\rho_{\rm DM} = \int_{0}^{\infty} {f_{\rm ST}(\sigma)MdM}\label{rdm},
\end{equation}

\noindent where $\rho_{\rm DM}$ is the comoving dark matter density.

We consider that the baryon distribution traces the dark matter distribution without bias. Thus, the density of baryons is proportional to the density of dark matter. The fact that  stars can form only in structures that are suitably dense can be parameterized by the threshold mass $M_{\rm min}$. Thus, the fraction of baryons at redshift $z$ that are in structures is given by

\begin{equation}
f_{\rm b}(z)=\frac{\int_{M_{\rm min}}^{M_{\rm max}} {f_{\rm ST}(\sigma)MdM}}{\int_{0}^{\infty} {f_{\rm ST}(\sigma)MdM}}.\label{fbaryon}
\end{equation}
 
With this definition, the baryon accretion rate $a_{\rm b}(t)$ which accounts for the increase in the fraction of baryons in structures is given by \citep{d1}:

\begin{equation}
a_{\rm b}(t) = \Omega_{\rm b}\rho_{\rm c}\left(\frac{dt}{dz}\right)^{-1}\left|\frac{df_{\rm b}(z)}{dz}\right|,\label{abaryon}
\end{equation}

\noindent where $\rho_{\rm c}=3H_{0}^{2}/8\pi G$ is the critical density of the Universe.

The age of the Universe that appears in (\ref{abaryon}) is related to the redshift by:
 
\begin{equation}
\frac{dt}{dz} = \frac{9.78h^{-1} \rm{Gyr}}{(1+z)\sqrt{\Omega_{\Lambda}+\Omega_{\rm m}(1+z)^{3}}},\label{timez}
\end{equation}

In Eq. (\ref{fbaryon}) we have used as upper limit $M_{\rm max}= 10^{18}\, M_{\odot}$. This choice permits a reasonable computational time to run the models. Moreover, models with $M_{\rm max}= 10^{24}\, M_{\odot}$ showed no considerable difference in the results. In the next Section, we discuss how to obtain the CSFR from the hierarchical scenario here described.

\section{The Cosmic Star Formation}

In hierarchical models for galaxy formation the first star-forming halos are predicted to collapse at redshift $z\gtrsim 20$, having masses $\sim 10^{6}{\rm M}_{\odot}$ \citep{sal}. In particular, the star formation history for a `galactic-like system' is determined by the interplay between incorporation of baryons into collapsed objetcs (stars, stellar remnants, and smaller objects) and return of baryons into diffuse state (gaseous clouds and intercloud medium of the system).

The later process can be two-fold: (a) mass return from stars to the `interstellar medium of the system' through, for example, stellar winds, and supernovae, which happens at the local level; and (b) net global infall of baryons from outside of the system. The former process is a well-known and firmly established part of the standard stellar evolution lore (see, e.g., \citealt{chiosi}), and although details of mass-loss in a particular stellar type may still be controversial, there is nothing controversial in the basic physics of this process.

Thus, we use throughout this paper the basic process above described. To do that, we consider the baryon accretion rate $a_{\rm b}(t)$, described by Eq. (\ref{abaryon}), as an infall term that supplies the reservoir represented by the halos. Therefore, the number of stars formed by unity of volume, mass and time is given by:

\begin{equation}
\frac{d^{3}N}{dVdmdt} = \Phi (m) \Psi(t),\label{d3n}
\end{equation}

\noindent where $\Phi (m)$ is the initial mass function (IMF) which gives the distribution function of stellar masses, and $\Psi (t)$ is the star formation rate. See that $\Psi (t)$ is assumed to be independent of mass while $\Phi (m)$ is assumed to be independent of time.

We use a Schmidt law \citep{sch1,sch2} for $\Psi (t)$. Therefore,

\begin{equation}
\frac{d^{2}M_{\star}}{dVdt} = \Psi (t) = k[\rho_{\rm g}(t)]^{\alpha},\label{sclaw}
\end{equation}

\noindent where $k$ is a constant that will be identified later, $\rho_{\rm g}$ is the local gas density, and $\alpha=1$. See that (\ref{sclaw}) shows that stars are formed by the gas contained in the halos.

On the other hand, we assume that the IMF follows the \citet{s1} form

\begin{equation}
\Phi(m) = A m^{-(1+x)}\label{imf1},
\end{equation}

\noindent where $x=1.35$ (our fiducial value) and $A$ is a normalization factor.

The constant $A$ is determined by the condition which all stars are formed into the mass range $[m_{\rm inf},m_{\rm sup}]$. That is,

\begin{equation}
\int_{m_{\rm inf}}^{m_{\rm sup}}Am^{-(1+x)}mdm = 1,\label{norimf}
\end{equation}

\noindent and we consider $m_{\rm inf}=0.1{\rm M}_{\odot}$ and $m_{\rm sup}=140{\rm M}_{\odot}$ as limits in (\ref{norimf}).

The mass ejected from stars, for example through winds and supernovae, is returned to the `interstellar medium of the system'. Thus, we have: 

\begin{equation}
\frac{d^{2} M_{\rm ej}}{dVdt} = \int_{m(t)}^{\rm M_{sup}}{(m-m_{\rm r})\Phi(m)\Psi(t-\tau_{m})dm},\label{mej1}
\end{equation}

\noindent where the lower limit of the integral, $m(t)$, corresponds to the stellar mass whose lifetime is equal to $t$. In the integrand, $m_{\rm r}$ is the mass of the remnant, which depends on the progenitor mass, and the star formation rate is taken at the retarded time $(t-\tau_{\rm m})$, where $\tau_{\rm m}$ is the lifetime of a star of mass $m$. 

For all stars formed in the halos, we use the metallicity-independent fit of \citet{s6,c4}

\begin{equation}
 \log_{10}(\tau_{\rm m})=10.0-3.6\,\log_{10}\left(\frac{M}{\rm M_{\odot}}\right) +\left[ \log_{10}
\left( \frac{M}{\rm M_{\odot}}\right) \right]^{2},
\end{equation}

\noindent where $\tau_{\rm m}$ is the stellar lifetime given in years.

The mass of the remnant, $m_{\rm r}$, in Eq. (\ref{mej1}) is calculated using the following assumptions:

\noindent a) Stars with $m < 1\ {\rm M}_{\odot}$ have a high lifetime so they do not contribute for $M_{\rm ej}$;

\noindent b) Stars with $1\ {\rm M}_{\odot} \leq m\leq 8\ {\rm M}_{\odot}$ after evolving off the main sequence left carbon-oxygen white dwarfs as remnants, where 

\begin{equation}
m_{\rm r} = 0.1156\ m +0.4551;
\end{equation}

\noindent c) Stars in the range $8\ {\rm M}_{\odot} < m\leq 10\ {\rm M}_{\odot}$ after evolving off the main sequence left oxygen-neon-magnesium white dwarfs with $m_{\rm r} = 1.35\ {\rm M}_{\odot}$;

\noindent d) Stars with $10\ {\rm M}_{\odot} < m< 25\ {\rm M}_{\odot}$ explode as supernovae leaving neutron stars as remnants ($m_{\rm r}=1.4\ {\rm M}_{\odot}$);

\noindent e) Stars with $25\ {\rm M}_{\odot} \leq m\leq 140\ {\rm M}_{\odot}$ produce black hole remnants. In this case, we consider that $m_{\rm r}=m_{\rm He}$. Note that $m_{\rm He}$ is the mass of the helium core before collapse (see \citealt{h1}). Thus,

\begin{equation}
m_{\rm r}=m_{\rm He}=\frac{13}{24}(m-20\ {\rm M}_{\sun}).\label{mr}
\end{equation} 

Then, using equations (\ref{sclaw}) and (\ref{mej1}) we can write an equation governing the total gas density ($\rho_{g}$) in the halos. Namely,

\begin{equation}
 \dot\rho_{\rm g}=-\frac{d^{2}M_{\star}}{dVdt}+\frac{d^{2}M_{\rm ej}}{dVdt}+a_{\rm b}(t)\label{rhogas},
\end{equation}

\noindent where $a_{\rm b}(t)$, Eq. (\ref{abaryon}), gives the rate at which the halos accrete mass.

Numerical integration of (\ref{rhogas}) produces the function $\rho_{\rm g}(t)$ at each time $t$ (or redshift $z$). Once obtained $\rho_{\rm g}(t)$, we return to Eq. (\ref{sclaw}) in order to obtain the `Cosmic Star Formation Rate' $\Psi(t)$. Just replacing $\Psi(t)$ by $\dot\rho_{\star}$ we can write

\begin{equation}
\dot\rho_{\star}=k\rho_{\rm g}\label{csfr},
\end{equation}

\noindent where the constant $k$ represents the inverse of the timescale for star formation. Namely, $k=1/\tau_{\rm s}$.

We normalize the CSFR in order to produce $\dot\rho_{\star}=0.016\,{\rm M}_{\odot}\,{\rm yr}^{-1}\,{\rm Mpc}^{-3}$ at $z=0$. With this normalization, we obtain a good agreement with both the present value of the CSFR derived by \citet{sprher}, who employed hydrodynamic simulations of structure formation, and the observational points taken from \cite{h2,h3}.

The cosmological parameters used in our models are: $\Omega_{\Lambda}= 0.76$, $\Omega_{\rm m} = 0.24$, $\Omega_{\rm b} = 0.04$, $\sigma_{8}=0.84$, and Hubble constant $H_{0}=100\,h\,{\rm km}\,{\rm s}^{-1}\,{\rm Mpc}^{-1}$ with $h=0.73$.

In the Figure 1 we present the CSFR derived from our models in function of the threshold mass $M_{\rm min}$ (see Eq. \ref{fbaryon}). We use a IMF with slope $x=1.35$, $\tau_{\rm s} = 2.0\ {\rm Gyr}$ as timescale for star formation, and we consider that stars start to form at redshift $z_{\rm ini}=20$. As can be seen, models with $M_{\rm min}=10^{6}-10^{8}{\rm M}_{\odot}$ have an excellent agreement with the observational CSFR at redshifts $z\lesssim 6.5$. See that the threshold mass $M_{\rm min}$ act on the amplitude and the redshift ($z_{\star}$) at which the amplitude of the CSFR is maximum.

The model with $M_{\rm min}=10^{10}{\rm M}_{\odot}$ has a good agreement with data at $z\lesssim 5$. On the other hand, at more higher redshifts ($5\lesssim z \lesssim 6.5$) this model does not agree very well with the observational points. In Figure 1 we also included the CSFR derived by Springel \& Hernquist (SH) for comparison. Although our models with $M_{\rm min}=10^{6}-10^{8}{\rm M}_{\odot}$ have an amplitude greater than that derived by SH we can observe that both, SH and our models, fit very well the observational data.

\begin{figure}
\includegraphics[width=90mm]{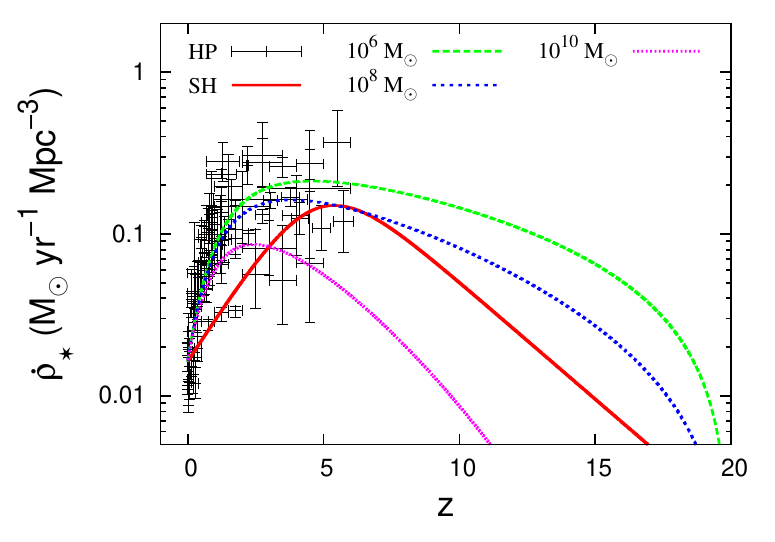}
\caption{The CSFR obtained from our models compared to the observational points (HP) taken from \citet{h2,h3}. We used a standard Salpeter IMF ($x=1.35$), and $\tau_{\rm s} = 2.0\ {\rm Gyr}$ as timescale for star formation. In this plot, we can see the influence of $M_{\rm min}$ (the threshold mass for halo formation) on the CSFR. The solid line represents the Springel \& Hernquist CSFR (SH), the dashed line corresponds to $M_{\rm min}=10^{6}{\rm M}_{\odot}$, the short dashed line corresponds to $M_{\rm min}=10^{8}{\rm M}_{\odot}$, and the dotted line represents $M_{\rm min}=10^{10}{\rm M}_{\odot}$.}
\end{figure}

In Figure 2 we show the influence of $\tau_{\rm s}$ on the CSFR. We consider $x=1.35$, $z_{\rm ini}=20$, and we take $M_{\rm min}=10^{6}{\rm M}_{\odot}$ for the threshold mass. Note that, $\tau_{\rm s} \leq 2.0\ {\rm Gyr}$ produces a gas comsuption timescale compatible with early type galaxies \citep{pach1}. Thus, the first basic effect of increasing $\tau_{s}$ is to shift the peak of the CSFR to lower redshifts. That means, the higher the $\tau_{\rm s}$ parameter, the lower is the readshift where appears the peak of $\dot\rho_{\star}$. In particular, the peak of $\dot\rho_{\star}$ is shifted from redshift 3.3 if $\tau_{\rm s} = 4.0\ {\rm Gyr}$ to 6.1 if $\tau_{\rm s} = 1.0\ {\rm Gyr}$.

The parameter $\tau_{\rm s}$ is also related to the amplitude  of $\dot\rho_{\star}$ (see also Eq. \ref{csfr}). See that considering $M_{\rm min}=10^{6}{\rm M}_{\odot}$ then the models with $\tau_{\rm s} = 2.0-3.0\ {\rm Gyr}$ are those that present the best concordance with the observational data. It is worth stressing that both parameters, $M_{\rm min}$ and $\tau_{\rm s}$, produce similar effects on the results. That is, they act on the amplitude of $\dot\rho_{\star}$ and on the value of $z_{\star}$. In Figure 2 is also included the CSFR derived by Springel \& Hernquist for comparison.

\begin{figure}
\includegraphics[width=90mm]{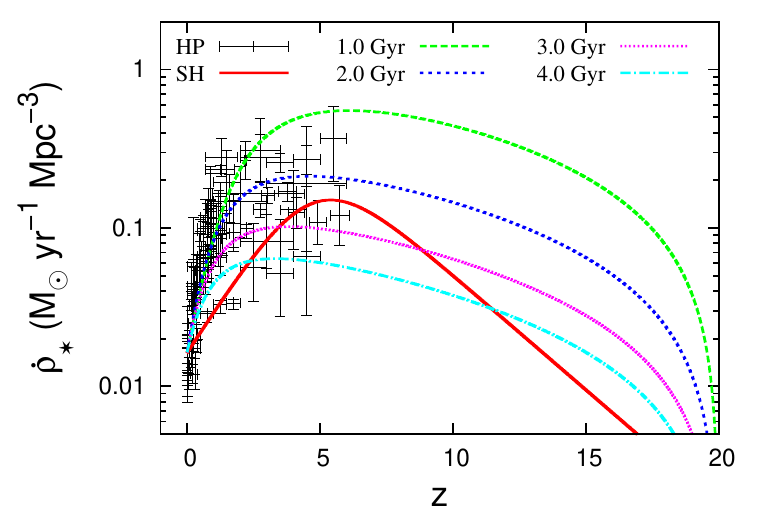}
\caption{The influence of the timescale for star formation ($\tau_{\rm s}$) on the results. The solid line represents the Springel \& Hernquist CSFR (SH), the dashed line corresponds to $\tau_{\rm s}=1.0\ {\rm Gyr}$, the short dashed line corresponds to $\tau_{\rm s}=2.0\ {\rm Gyr}$, the dotted line corresponds to $\tau_{\rm s}=3.0\ {\rm Gyr}$, and the dot-dashed line represents $\tau_{\rm s}=4.0\ {\rm Gyr}$. These models have a threshold mass $M_{\rm min} = 10^{6}{\rm M}_{\odot}$, and a IMF with slope $x=1.35$. HP stands for the observational CSFR \citep{h2,h3}.}
\end{figure}

In Figure 3 we see the influence of $z_{\rm ini}$ on the evolution of $\dot\rho_{\star}$. The models have similar evolution at $z\lesssim 5$. However, at larger redshifts, the model with $z_{\rm ini}=40$ produces a CSFR higher than that obtained from $z_{\rm ini}=20$. In particular, the peak of the CSFR occurs at redshift 4.6 (5.5) for the model with $z_{\rm ini}=20$ ($z_{\rm ini}=40$).

\begin{figure}
\includegraphics[width=90mm]{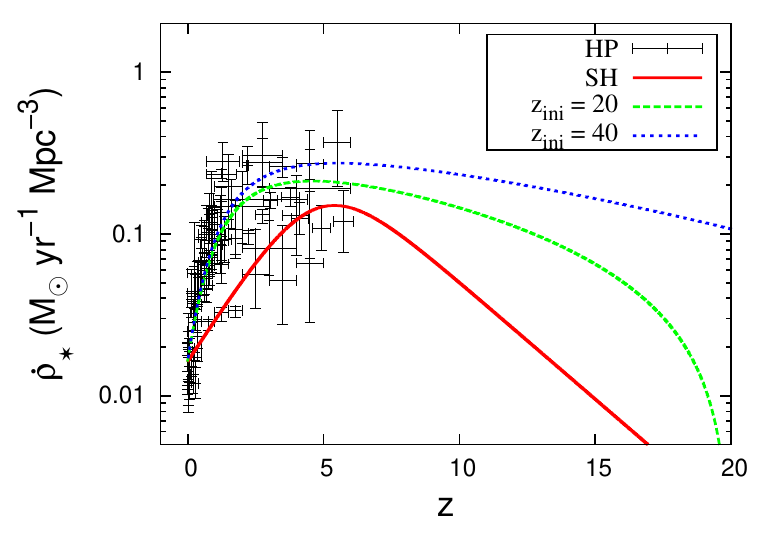}
\caption{Models with $\tau_{\rm s}=2.0\ {\rm Gyr}$ and $M_{\rm min} = 10^{6}{\rm M}_{\odot}$ but considering two different values for the initial redshift. The solid line corresponds to the Springel \& Hernquist CSFR (SH), the dashed line corresponds to $z_{\rm ini}=20$, and the short dashed line corresponds to $z_{\rm ini}=40$. HP stands for the observational CSFR \citep{h2,h3}.}
\end{figure}

It is worth stressing that the CSFR is inferred from observations of the light emitted by stars at various wavelengths. These observable samples are flux-limited, and thus the intrinsic luminosity of the faintest objects in the sample changes with redshift. This incompleteness of the samples is corrected by using a functional (Schechter function) to the luminosity function obtained from the observations themselves.

An important parameter on the determination of the CSFR is the obscuration by dust that is well known to affect measurements of galaxy luminosty at ultraviolet (UV) and optical wavelengths. Correcting for this effect is not always straightforward. Thus, there are large uncertainties associated to the determination of the CSFR as can be seen from Figures $1-3$ (see, in particular, \citealp{h2,d2} who discuss these uncertainties with more details).

\section[]{The stochastic background of gravitational waves}

In this section we use the CSFR ($\dot\rho_{\star}$) obtained from the hierarchical model to determine the stochastic background of gravitational waves (SBGWs) generated by stars which collapse to black holes. Initially, we present a quick overview on the formalism used to characterize a SBGWs because this subject is discussed in previous works (see, for example, \citealp{d3,d4,d5,d2,m1}). After this quick overview we display and compare the results of the models considered.

Let us write the specific flux received in GWs at the present epoch as

\begin{equation}
F_{\nu}(\nu_{\rm obs})= \int \frac{l_{\nu}}{4\pi d_{\rm L}^{2}}
\frac{d\nu}{d\nu_{\rm obs}}dV,
\end{equation}

\noindent where

\begin{equation}
l_{\nu} = \frac{dL_{\nu}}{dV}
\end{equation}

\noindent is the comoving specific luminosity density (given, e.g, in ${\rm erg\, s^{-1}\, Hz^{-1}\, Mpc^{-3}}$), which obviously refers to the source frame. See that $dV$ is the comoving volume element, and $d_{\rm L}$ is the luminosity distance.

The above equations are valid to estimate a stochastic background radiation received on Earth independent of its origin. In the present paper $l_{\nu}$ can be written as follows

\begin{equation}
l_{\nu} = \int \frac{dE_{\rm GW}}{d\nu}\dot\rho_{\star}(z)
\Phi(m)dm,
\end{equation}

\noindent where $dE_{\rm GW}/d\nu$ is the specific energy of the source. Note that in the above equation $\dot\rho_{\star}(z)$
is the CSFR, and $\Phi(m)$ is the IMF.

Thus, the flux $F_{\nu}(\nu_{\rm obs})$ received on Earth reads

\begin{equation}
F_{\nu}(\nu_{\rm obs})= \int \frac{1}{4\pi d_{\rm L}^{2}}
\frac{dE_{\rm GW}}{d\nu}\frac{d\nu}{d\nu_{\rm
obs}}\dot\rho_{\star}(z) \Phi(m)dmdV.
\end{equation}

In particular, one can write the differential rate of production of GWs, for the case of a background produced by an ensamble of black holes, as follows

\begin{equation}
dR_{\rm BH} = \dot\rho_{\star}\frac{dV}{dz}\Phi(m)dmdz\label{dR_BH}.
\end{equation}

Using Eq.(\ref{dR_BH}) it follows that

\begin{equation}
F_{\nu}(\nu_{\rm obs})= \int \frac{1}{4\pi d_{\rm L}^{2}}\frac{dE_{\rm GW}}{d\nu} \frac{d\nu}{d\nu_{\rm obs}}
dR_{\rm BH}.
\end{equation}

Note that in the above equation, what multiplies $dR_{\rm BH}$ is nothing but the specific energy flux per unity frequency (in, e.g., ${\rm erg\, cm^{-2}\, Hz^{-1}}$), i.e.,

\begin{equation}
f_{\nu}(\nu_{\rm obs}) = \frac{1}{4\pi d_{\rm L}^{2}}\frac{dE_{\rm
GW}}{d\nu} \frac{d\nu}{d\nu_{\rm obs}}.
\label{flux01}
\end{equation}

On the other hand, the specific energy flux per unit frequency for GWs is given by \citep{carr}

\begin{equation}
f_{\nu}(\nu_{\rm obs}) = \frac{\pi c^{3}}{2G}h_{\rm BH}^{2}.
\label{flux02}
\end{equation}

Also, the spectral energy density, the flux of GWs, received on
Earth, $F_{\nu}$, in ${\rm {erg}\,{cm}^{-2}\,{s}^{-1}\,{Hz}^{-1}}$ can be written as

\begin{equation}
F_{\nu}(\nu_{\rm obs})= \frac{\pi c^{3}}{2G}h_{\rm BG}^{2}\nu_{\rm
obs}.
\end{equation}

From the above equations one obtains

\begin{equation}
h_{\rm BG}^{2} = \frac{1}{\nu_{\rm obs}}\int h_{\rm BH}^{2}
dR_{\rm BH}.
\end{equation}

See that $h_{\rm BH}$ is the dimensionaless amplitude produced by the collapse of a star to form a black hole. Its expression is obtained from \citet{t1}. Thus,

\begin{equation}
h_{\rm BH} \simeq 7.4 \times 10^{-20} \epsilon_{\rm GW}^{1/2}\left(\frac{m_{\rm r}}{\rm M_{\sun}}\right)\left(\frac{d_{\rm L}}{1\rm{Mpc}}\right)^{-1},\label{h_BH}
\end{equation}

\noindent where $\epsilon_{\rm GW}$ is the efficiency of generation of GW's, and $m_{\rm r}$ is the mass of the black hole formed.

It is worth mentioning that Eq. (\ref{h_BH}) refers to the black hole `ringing', which has to do with the de-excitation of the black hole quasi-normal modes.

The collapse of a star to black hole produces a signal with frequency $\nu_{\rm obs}$ given by

\begin{equation}
\nu_{\rm obs} \simeq 1.3 \times 10^{4} {\rm Hz}\left(\frac{\rm{M}_{\sun}}{m_{\rm r}}\right)(1+z)^{-1},
\end{equation}

\noindent where the factor $(1+z)^{-1}$ takes into account the redshift effect on the emission frequency. That is, a signal emitted at frequency $\nu_{\rm e}$ at redshift $z$ is observed at frequency $\nu_{\rm obs} = \nu_{\rm e} (1+z)^{-1}$.

As discussed in the previous section, we consider that black holes are formed from stars with $25\ {\rm M}_{\odot} \leq m \leq 140\ {\rm M}_{\odot}$. The mass of the remnant is taken to be the mass of the helium core before collapse (see Eq. \ref{mr}). 

Another relevant physical quantity associated with the SBGWs is the closure energy density per logarithmic frequency span, which is given by

\begin{equation}
\Omega_{\rm GW}=\frac{1}{\rho_{\rm c}}\frac{{\rm d}\ \rho_{\rm GW}}{{\rm d}\ \log{\nu_{\rm obs}}}.
\end{equation}

The above equation can be re-written as

\begin{equation}
\Omega_{\rm GW}=\frac{\nu_{\rm obs}}{c^{3}\rho_{\rm c}}F_{\nu}=\frac{4\pi^{2}}{3H_{0}^{2}}\nu_{\rm obs}^{2} h_{\rm BG}^{2}.
\end{equation}

Thus, given a star formation history, consisting of a star formation rate per comoving volume (CSFR), $\dot\rho_{\star}(z)$, and an initial mass function (IMF), $\Phi(m)$, the stochastic background of gravitational waves produced by pre-galactic black holes can be characterize.

Finally, to assess the detectability of a GW signal, one must evaluate the signal-to-noise ratio (S/N), which for a pair of interferometers is given by (see, for example, \citealp{crs,f5,a1,d4,d5,regimbau})

\begin{equation}
{\rm (S/N)^{2}} = \left[\left({\frac{9H_{0}^{4}}{50\pi^{4}}}\right) T\int_{0}^{\infty}d\nu\frac{\gamma^{2}(\nu)\Omega_{\rm GW}^{2}(\nu)}{\nu^{6}S_{\rm h}^{(1)}(\nu)S_{\rm h}^{(2)}(\nu)}\right],\label{signal}
\end{equation}

where $S_{\rm h}^{(i)}$ is the spectral noise density, $T$ is the integration time, and $\gamma(\nu)$ is the overlap reduction function, which depends on the relative positions and orientations of the two interferometers. For the $\gamma(\nu)$ function we refer the reader to \citet{f5} who was the first to calculate a closed form for the LIGO observatories.

Using the formalism described above and in the previous sections we study a total number of 72 models varying the following parameters:

\noindent a) the threshold mass ($M_{\rm min}$) for structure formation, where we consider the values $ 10^{6}{\rm M}_{\odot}$, $10^{8}{\rm M}_{\odot}$, and $10^{10}{\rm M}_{\odot}$;

\noindent b) the exponent ($x$) of the IMF, where we consider $x=1.35$ (`Salpeter exponent'), $x=0.35$ which yields a higher number of black hole remnants than Salpeter IMF, and $x=2.35$ which produces a lower number of black hole remnants than Salpeter exponent;

\noindent c) the timescale for star formation ($\tau_{\rm s}$), where we consider the values 1.0 Gyr, 2.0 Gyr, 3.0 Gyr, and 4.0 Gyr;

\noindent d) the initial redshift ($z_{\rm ini}$) where star formation begins to occur. We take the values 20 and 40.

On this set of models we use two criteria for selecting the best ones. The first criterion is to have good agreement with observational star formation data at redshifts $z\lesssim 6.5$ \footnote{We performed $\chi^{2}$ analysis over the models with ${\rm (S/N)} > 3$. In particular, we determine the reduced chi-square defined as $\chi_{\rm r} = \chi^{2}/{\rm dof}$ (where ``dof" means ``degrees of freedom"). We consider that models with $\chi_{\rm r}\leq 1$ have good agreement with observational data.}. The second criterion is to produce a signal-to-noise ${\rm (S/N)} > 3$ for a pair of `advanced' interferometers. We consider this choice of (S/N) as reasonable for an adequate characterization of the SBGWs.

Figure 4 presents the models with $M_{\rm min}=10^{6}{\rm M}_{\odot}$ which satisfy the above criteria.

\begin{figure}
\includegraphics[width=90mm]{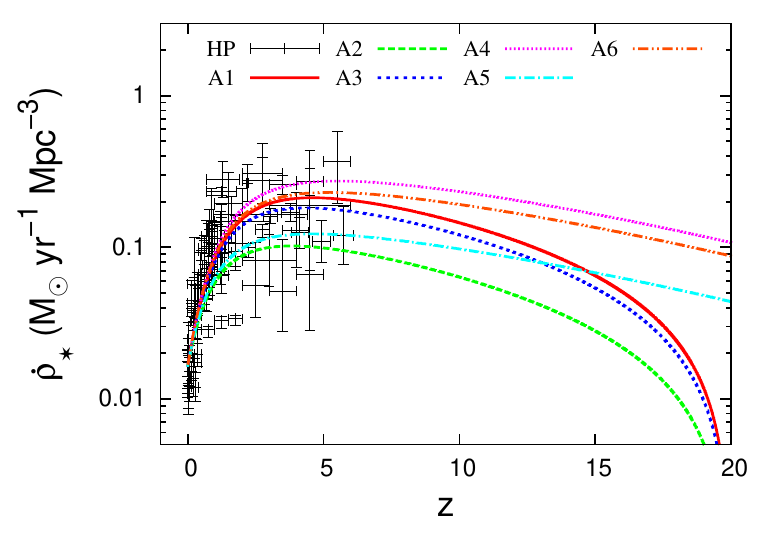}
\caption{The CSFR for models with $M_{\rm min}=10^{6}{\rm M}_{\odot}$ and good agreement with observational data. The main characteristics of these models are described in Table 1.}
\end{figure}

Table 1 shows the main results for the six models ${\rm A1-A6}$ which are presented in Figure 4. The efficiency of generation of GWs  is taken from \citet{sp86} who simulated the axisymmetric collapse of a rotating star to black hole. We use their maximum value, namely, $\epsilon_{\rm GW_{max}}=7\times 10^{-4}$. We will discuss below the dependence of $\epsilon_{\rm GW}$ on the results.

\begin{table}
\caption{The main results of the models with $M_{\rm min}=10^{6}{\rm M}_\odot$ of the Figure 4. The signal-to-noise, (S/N), is presented for a pair of LIGO III (advanced configuration) interferometers. (S/N) is computed for one year of observation and we consider a gravitational wave efficiency $\epsilon_{\rm GW_{max}} = 7 \times 10^{-4}$.}
 \label{tab1}
 \begin{tabular}{@{}lccccc}
  \hline
 Model & $z_{\rm ini}$ & $x\ (\rm IMF)$ & $\tau_{\rm s}\ {\rm Gyr}$ & $z_{\star}$  & (S/N) \\
 \hline
 A1  &  20  &  1.35  &  2.0  &  4.6  &  7.4    \\
 A2  &  20  &  1.35  &  3.0  &  3.8  &  3.8    \\
 A3  &  20  &  0.35  &  1.0  &  4.4  &  93.5   \\
 A4  &  40  &  1.35  &  2.0  &  5.6  &  9.8    \\
 A5  &  40  &  1.35  &  3.0  &  4.6  &  4.8    \\
 A6  &  40  &  0.35  &  1.0  &  5.3  &  119.9  \\
 \hline
\end{tabular}

\medskip

\end{table}

See that to calculate de signal-to-noise ratio we consider that the integration time in Eq. (\ref{signal}) is one year. In the fifth column of Table 1 we present the redshift ($z_{\star}$) where the CSFR reaches its maximum value; in the sixth column we present the signal-to-noise ratio (S/N).

Note that there is possibility of detecting the SBGWs here proposed if $\epsilon_{\rm GW}$  is around the maximum value. Observe that for Salpeter IMF ($x=1.35$) we obtain a significant (S/N) if $\tau_{\rm s} \sim 2.0-3.0\ {\rm Gyr}$.

On the other hand, the models with $x=0.35$ produce the highest values for the (S/N). This happens because $x=0.35$ produces a higher number of massive stars than the Salpeter IMF. In this case, the CSFR that fit the observational data are those with $\tau_{\rm s}\lesssim 1.0\ {\rm Gyr}$. See that the models with $\tau_{\rm s}\lesssim 1.0\ {\rm Gyr}$ have a short timescale for star formation. These values for the parameter $\tau_{\rm s}$ are consistent with a high-mass stellar population.

However, if the IMF of pre-galactic stars is close to $x=2.35$ then there is no hope of detecting the SBGWs we proposed here, even for ideal orientation and locations of the LIGO interferometers. In particular, all models with $x=2.35$ have ${\rm (S/N)}<0.1$, same for those models producing $\dot\rho_{\star}$ with excellent agreement with Hopkins data. Thus, the first conclusion is that it would be possible the detection of a background of pre-galactic black holes if the IMF of these objects is $x \gtrsim 1.35$ and if $\epsilon_{\rm GW_{max}} \sim 7 \times 10^{-4}$.

In order to see the influence of $M_{\rm min}$ on the value of the signal-to-noise ratio we present in Figure 5 the models with $M_{\rm min}=10^{8}{\rm M}_\odot$ and that satisfy our two criteria as above defined.

\begin{figure}
\includegraphics[width=90mm]{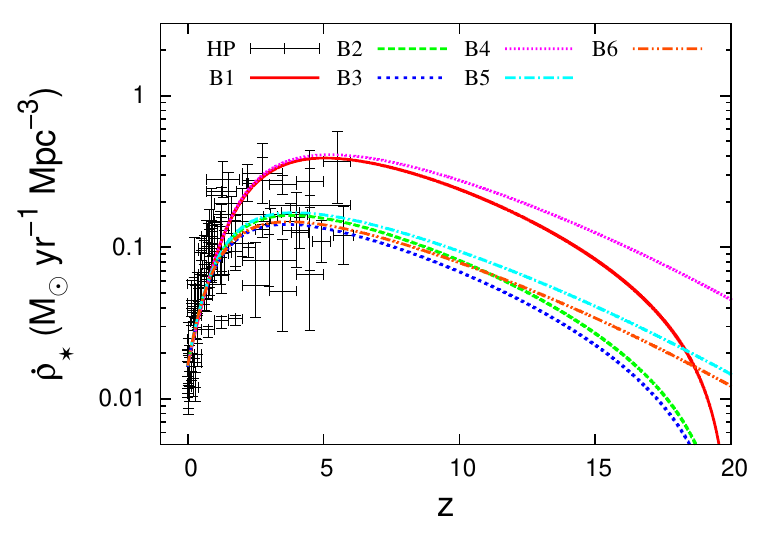}
\caption{The CSFR for models with $M_{\rm min}=10^{8}{\rm M}_{\odot}$ and good agreement with observational data. The main characteristics of these models are described in Table 2.}
\end{figure}

Table 2 shows the main results for the six models ${\rm B1-B6}$ which are presented in Figure 5. The first effect of $M_{\rm min}$ is to shift $z_{\star}$ (for example, compare models A1 and B2). That is, a halo with mass $10^{6}{\rm M}_{\odot}$ collapses earlier than a halo with mass $10^{8}{\rm M}_{\odot}$. Thus, the maximum of star formation for models with $10^{8}{\rm M}_{\odot}$ will be shifted to low redshifts.

The second effect is on the amplitude of $\dot\rho_{\star}$ as discussed in the previous section. As the quantity of black holes is $\propto\dot\rho_{\star}$ then increasing the value of $M_{\rm min}$ the number of black holes formed will decrease. As a consequence, models with $10^{8}{\rm M}_{\odot}$ present a lower (S/N) than those with $M_{\rm min}=10^{6}{\rm M}_{\odot}$.

The third effect can be seen comparing Tables 1 and 2. The models which satisfy the selection criteria with $M_{\rm min}=10^{6}{\rm M}_\odot$ are those with $\tau_{\rm s} \sim 2.0-3.0\,{\rm Gyr}$ for $x=1.35$. Otherwise, with $M_{\rm min}=10^{8}{\rm M}_\odot$ the selection criteria are satisfied if $\tau_{\rm s} \sim 1.0- 2.0\,{\rm Gyr}$ for $x=1.35$. This result can be understood remembering that $\tau_{\rm s}$ also acts on the amplitude of $\dot\rho_{\star}$.

That means, if we decrease the value of $\tau_{\rm s}$ the amplitude of $\dot\rho_{\star}$ increases (see, for an instance, Figure 2 and Eq. \ref{csfr}). On the other hand, as above discussed, if we increase the parameter $M_{\rm min}$, the amplitude of $\dot\rho_{\star}$ is reduced. Thus, if we change $M_{\rm min}$ from $10^{6}\, {\rm M}_\odot$ to $10^{8}\,{\rm M}_\odot$, we have to decrease the parameter $\tau_{\rm s}$ in order to obtain $\dot\rho_{\star}$ with good agreement with the observational data and also to produce ${\rm (S/N)}>3$.

\begin{table}
\caption{The main results of the models with $M_{\rm min}=10^{8}{\rm M}_\odot$.}
 \label{tab2}
 \begin{tabular}{@{}lccccc}
  \hline
 Model & $z_{\rm ini}$ & $x\ (\rm IMF)$ & $\tau_{\rm s}\ {\rm Gyr}$ & $z_{\star}$  & (S/N) \\
 \hline
 B1  &  20  &  1.35  &  1.0  &  5.1  &  11.9  \\
 B2  &  20  &  1.35  &  2.0  &  3.8  &  5.7   \\
 B3  &  20  &  0.35  &  1.0  &  3.6  &  72.8   \\
 B4  &  40  &  1.35  &  1.0  &  5.2  &  13.2  \\
 B5  &  40  &  1.35  &  2.0  &  3.9  &  6.2   \\
 B6  &  40  &  0.35  &  1.0  &  3.8  &  77.6  \\
 \hline
\end{tabular}

\medskip

\end{table}

In Figure 6 we present the models with $M_{\rm min} = 10^{10}{\rm M}_{\odot}$ (see details of the models in Table 3). Only those with $x=1.35$ and $\tau_{\rm s} = 1.0\,{\rm Gyr}$ have a good agreement with observational data and produce ${\rm (S/N)}>3$. See that the difference between models C1 and C2 is very subtle. 

This happens because the fraction of baryons in strcutures with $M> 10^{10}{\rm M}_\odot$ is very small at redshifts $20-40$. Thus, $z_{\rm ini}$ does not have strong influence on the evolution of the models C1 and C2 at low redshifts.

\begin{figure}
\includegraphics[width=90mm]{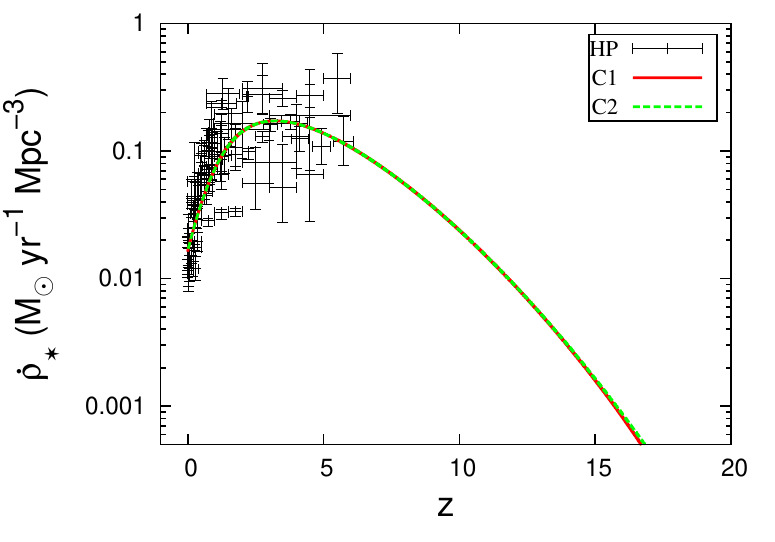}
\caption{The CSFR for models with $M_{\rm min}=10^{10}{\rm M}_{\odot}$ and good agreement with observational data. The main characteristics of these models are described in Table 3.}
\end{figure}

\begin{table}
\caption{The main results of the models with $M_{\rm min}=10^{10}{\rm M}_\odot$.}
 \label{tab3}
 \begin{tabular}{@{}lccccc}
  \hline
 Model & $z_{\rm ini}$ & $x\ (\rm IMF)$ & $\tau_{\rm s}\ {\rm Gyr}$ & $z_{\star}$  & (S/N) \\
 \hline
 C1  &  20  &  1.35  &  1.0  &  3.2  &  5.4   \\
 C2  &  40  &  1.35  &  1.0  &  3.2  &  5.6   \\
 \hline
\end{tabular}

\medskip

\end{table}

Figure 7 shows the density parameter $\Omega_{\rm GW}$ as a function of the observed frequency $\nu_{\rm obs}$. The density parameter increases at low frequencies and it reaches a maximum amplitude of about $9.0\times 10^{-7}$ around $200\ {\rm Hz}$ in the model A6. On the other hand, model A2 \footnote{The model A2 is that which has the smallest values for $\Omega_{\rm GW}$. As a consequence, from all models presented in Tables $1-3$, A2 is that which present the smallest signal-to-noise ratio.} reaches a maximum amplitude of $4.2\times 10^{-8}$ also around $200\ {\rm Hz}$. See that both the maximum amplitude of $\Omega_{\rm GW}$ and the high-frequency part of the spectra \footnote{Concerning for the results presented in Figures 7 and 8 we are defining the high-frequency part of the spectra as that for which $\nu_{\rm obs}> 200\, {\rm Hz}$.} are not strongly dependent on the initial redshift $z_{\rm ini}$. To verify that, compare the models A1 and A4; A2 and A5; A3 and A6.

However, the value of $z_{\rm ini}$ has influence over the low-frequency part of the spectra as can be seen from Figure 7. This part of the spectrum is dominated by the population of black holes formed at redshifts $z\gtrsim 7$.

It is worth stressing that \citet{d5} assuming a Springel and Hernquist \citep{sprher} model of star formation obtained a similar result for $\Omega_{\rm GW}$. Their spectrum peaks at $\Omega_{\rm GW}\ h^{2}\approx 5\times 10^{-9}$ at $\nu_{\rm obs}\approx 200\ {\rm Hz}$ for a Salpeter IMF. Using $h=0.73$ we find $\Omega_{\rm GW}\sim 9\times 10^{-9}$ for their fiducial model.

This is a factor $\sim 5$ lower than the maximum amplitude of $\Omega_{\rm GW}$ obtained by our model A2. However, note that $\dot\rho_{\star}$ obtained from `model 3.0 Gyr' in Figure 2, which corresponds to model A2 in Table 1, is smaller than the Springel and Hernquist CSFR only in the range $4.5 \lesssim z\lesssim 8.2$. Thus, except for this interval in redshift, the rate of core collapse obtained from Springel and Hernquist CSFR is actually smaller than that obtained from model A2.

The cusp in the curves shown in the Figure 7 is produced by our choice to the energy flux (see Equations \ref{flux01} and \ref{flux02}). See that the closure energy density ($\Omega_{\rm GW}$) is directly proportional to the energy flux, and therefore more sensitive to its frequency dependence. Here, the specific energy flux is obtained from Equation (\ref{h_BH}), which takes into account the most relevant quasi-normal modes of a rotating black hole.

In particular, we refer the reader to \citet{d3} who discuss the formulation presented here and compare it to that used by \citet{f2} where the energy flux is a function of frequency. Thus, their closure energy density is broader than we use here. As a consequence, $\Omega_{\rm GW}$ obtained by \citet{f2} has a smoother peak than ours. However, as discussed in \citet{d3}, both formulations presented similar results.

Since some authors use, instead of $\Omega_{\rm GW}$, the gravitational strain $S_{\rm h}^{1/2}$, defined by \citet{a2} as

\begin{equation}
S_{\rm h}=\frac{3H_{0}^{2}}{4\pi^{2}}\frac{1}{\nu_{\rm obs}}\Omega_{\rm GW},
\end{equation}

\noindent we show this quantity in Figure 8.

\begin{figure}
\includegraphics[width=90mm]{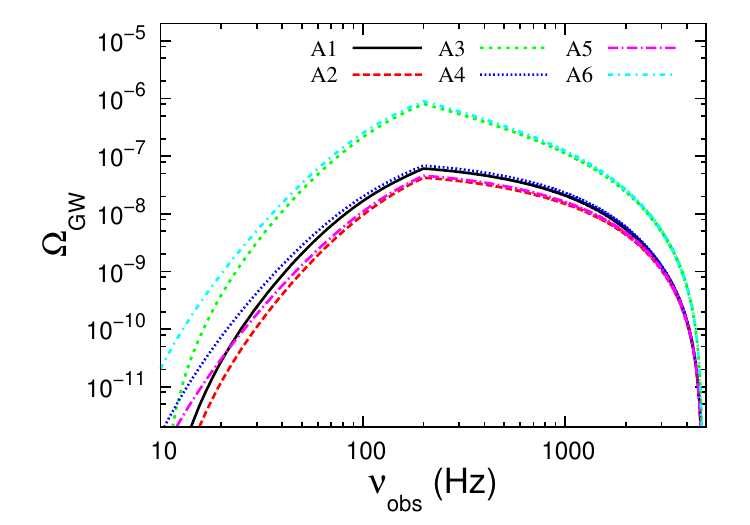}
\caption{Spectrum of the gravitational energy density parameter $\Omega_{\rm GW}$. Results are shown for the models ${\rm A1}-{\rm A6}$ of the Table 1.}
\end{figure}

\begin{figure}
\includegraphics[width=90mm]{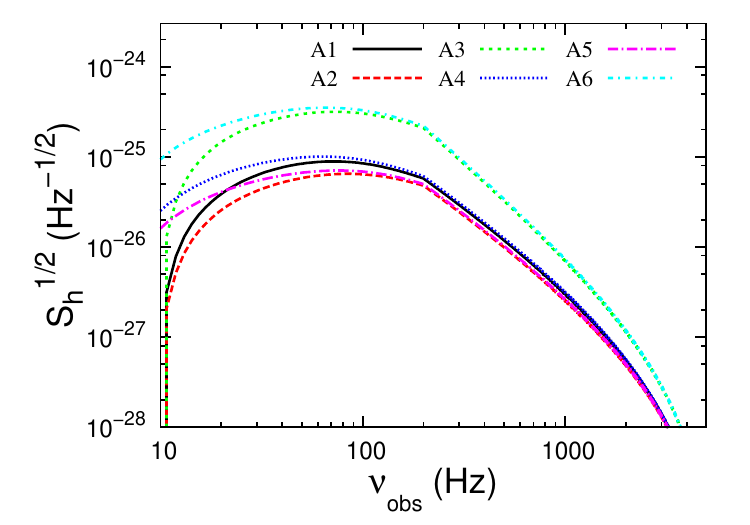}
\caption{Gravitational strain in ${\rm Hz}^{-1/2}$. Results are shown for the models ${\rm A1}-{\rm A6}$ of the Table 1.}
\end{figure}

A key parameter to determine the values presented in Tables 1, 2, and 3 is the efficiency of generation of GWs. We take the maximum efficiency found by \citet{sp86}, namely, $\epsilon_{\rm GW_{max}}=7\times 10^{-4}$ for an axisymmetric collapse resulting in a black hole.

On the other hand, more recently, \citet{f3} obtained the efficiency of $2\times 10^{-5}$ for a $100\ {\rm M}_{\odot}$ black hole remnant. Note that since $\Omega_{\rm GW} \propto \epsilon$, if the efficiency is actually closer to $2\times 10^{-5}$, the observed energy density in gravitational waves may be divided by a factor of 35. In this case, of all models here studied only model A6 will produce ${\rm (S/N)}>3$.

However, the distribution of $\epsilon_{\rm GW}$ in function of the mass of a black hole is unknown. In particular, let us think of what occurs with other compact objects $-$ namely, the neutron stars $-$ to see if we can learn something from them. A newly born neutron star could lose angular momentum due to gravitational waves associated with non-radial oscillations \citep{ferr03}. This could explain why all known young neutron stars are relatively slow rotators.

The black holes could have had a similar history, i.e, they could have been formed rapidly rotating and lost momentum to gravitational radiation via their quasi-normal modes. If this was the case, the value of $\epsilon_{\rm GW}$ could be near the maximum one, or in the worst case, it could have a value to produce ${\rm (S/N)} > 3$ for a LIGO III pair.

\section{Summary and Discussion}

In this work, we have used the hierarchical formation scenario derived from the Press-Schechter formalism to build the cosmic star formation rate - CSFR in a self-consistent way. Our paper differs from earlier works basically in the form as is obtained the function $\dot\rho_{\star}$ (or CSFR).

In particular, from the hierarchical scenario we obtain the baryon accretion rate, $a_{\rm b}(t)$, that supplies the gaseous reservoir in the halos. Thus, the term $a_{\rm b}(t)$ is treated as an infall term in our model.

This scenario is in agreement with the cold dark matter model of cosmological structure formation, where the first sources of light are expected to form in $\sim 10^{6}{\rm M}_{\odot}$ dark matter potential at $z\geq 20$.

Using $\dot\rho_{\star}$ we calculate the stochastic background of gravitational waves produced by pre-galactic black holes. We show that a significant amount of GWs is produced related to the history of CSFR studied here, and this SBGWs can in principle be detected by a pair of LIGO III interferometers. 

Note that signal-to-noise ratios ${\rm (S/N)}\sim 90$ could be obtained if the efficiency of generation of GWs is close to the maximum value ($\epsilon_{\rm GW_{max}}=7\times 10^{-4}$), if the IMF produces a high number of massive remnants ($x=0.35$), and if $z_{\rm ini}\sim 20$. Considering a Salpeter IMF ($x=1.35$), we obtain signal-to-noise ratios ${\rm (S/N)}\sim 10$.

The critical parameter to be constrained in the case of a non-detection is $\epsilon_{\rm GW}$. A non-detection would mean that the efficiency of GWs during the formation of black holes is not high enough. In reality, $\epsilon_{\rm GW_{max}}$ should be divided by a factor $> 35$ in the case of a non-detection.

It is worth mentioning that an IMF with $x=2.35$ could also be responsible for a non-detection same with $\epsilon_{\rm GW}=\epsilon_{\rm GW_{max}}$. However, $x=2.35$ produces a high number of low mass stars that is not in agreement with recent numerical simulations of the collapse and fragmentation of primordial clouds (see, e.g., \citealp{abel}).

Another possibility for a non-detection is that the pre-galactic stars are such that the black holes formed had masses $>500{\rm M}_{\odot}$. In this case, the GW frequency band would be out of the LIGO bandwidth. 

However, considering black holes formed from stars with masses $25\ {\rm M}_{\odot}\lesssim m \lesssim 140\ {\rm M}_{\odot}$, then the sensitivity of the future third generation of detectors could be high enough to increase one order of magnitude in the expected value of (S/N). Examples of such detectors are the Large Scale Cryogenic Gravitational Wave Telescope (LCGT) and the European antenna EGO (see \citealp{regimbau} and the references therein for a short discussion on this subject).

Specifically, around $650\ {\rm Hz}$ the planned strain noise for EGO will be a factor of $\sim 4$ higher than that provided for advanced LIGO configuration. This could represent a gain of a factor $\sim 5-20$ for the value of (S/N) considering two interferometers located at the same place (see \citealp{regimbau}). Thus, some models in Tables $1-3$ could survive with ${\rm (S/N)}>3$ same with $\epsilon_{\rm GW} \sim 2\times 10^{-5}$.

In particular, the detection of a background with significant (S/N) would permit us to obtain the curve $S_{\rm h}^{1/2}$ (or $\Omega_{\rm GW}$) versus $\nu_{\rm obs}$. From it, one can constrain $\dot\rho_{\star}$ at high redshifts and the gravitational wave efficiency ($\epsilon_{\rm GW}$). Thus, the detection and characterization of a SBGWs could be used as a tool for study of the star formation at high redshifts.

It is worth stressing that several astrophysical sources can contribute to the background of gravitational waves, as mentioned in the Introduction. In principle, it should be possible to distinguish different sources from the detected gravitational wave spectrum. That is, from the caractheristics of the observed curve $\Omega_{\rm GW}$ versus $\nu_{\rm obs}$.

For example, in the present work we have shown that cosmological stellar black holes ($3\lesssim {\rm M_{BH}}/{\rm M}_{\odot}\lesssim 65$), formed at $z_{\rm ini}\lesssim 20-40$, produce a stochastic background in the frequency range $\sim 10\,{\rm Hz}-5\,{\rm kHz}$. In particular, the gravitational wave spectra peak at $\nu_{\rm obs}\approx 200\,{\rm Hz}$. If the black hole Population forms at low redshifts (for example, $z_{\rm ini}\lesssim 10$), both the frequency where $\Omega_{\rm GW}$ peaks and the minimum frequency of the spectra will be shifted to greater frequencies than those presented here.

However, the shape of $\Omega_{\rm GW}$ does not considerably change if we consider the same gravitational wave energy power spectrum for the sources. On the other hand, more massive stars ($m> 200\,{\rm M}_\odot$) will shifted the peak of the spectra for low frequencies. See for a moment the results of \citet{msf09} for black hole remnants of Population III stars with masses $100-500\, {\rm M}_\odot$. Their spectrum peaks at $\nu_{\rm obs}=2.74\,{\rm Hz}$ ($\Omega_{\rm GW}\approx 5\times 10^{-15}$) and the maximum frequency of the background is $\sim 600\,{\rm Hz}$. 

Another example can be seen from the work of \citet{buo05}. The authors studied the gravitational wave background from all cosmic supernovae. Their fiducial model peaks at $\nu_{\rm obs}=6\,{\rm Hz}$ ($\Omega_{\rm GW}\approx 10^{-13}$) while the maximum frequency of the background is $\sim 3\,{\rm kHz}$ and the spectrum can extent to very low frequencies ($\nu_{\rm obs} \lesssim 10^{-4}{\rm Hz}$). Thus, in principle, it would be possible to identify the signatures of different backgrounds if we have the curve $\Omega_{\rm GW}$ versus $\nu_{\rm obs}$ over a large range in frequency.

Last but not least, we refer the reader to the work of \citet{kh00} who present a unified model for the evolution of galaxies and quasars. Specifically, these authors discuss that gas cooling is not efficient in too massive structures and so haloes with circular velocity greater than $600\, {\rm km\,s^{-1}}$ could not form stars. If we take into account their results then the upper limit, $M_{\rm max}$, in Equation (\ref{fbaryon}) should be changed for $\sim 10^{13}{\rm M}_{\odot}$.

We checked all the models described in Tables $1-3$ with this new upper limit ($M_{\rm max}=10^{13}{\rm M}_{\odot}$). We verify that the amplitude of the CSFR decreases slightly at $z\lesssim 3.5$ when compared with the results obtained using $M_{\rm max}=10^{18}{\rm M}_{\odot}$ (at $z> 3.5$ we do not observe any modification in the behaviour of $\dot\rho_{\star}$). For the models with $M_{\rm min}=10^{6}{\rm M}_{\odot}$ ($10^{8}{\rm M}_{\odot}$) there is only a subtle modification in the final results. In particular, the signal-to-noise ratios are $\sim 3.9\%$ ($\sim 4.6\%$) lower than those presented in Table 1 (2). For the models with $M_{\rm min}=10^{10}{\rm M}_{\odot}$ we note a modification $\sim 8.6\%$ in the results of the Table 3. However, all models presented in Tables $1-3$ satisfy the ``two criteria'', as discussed in Section 4. That is, same using $M_{\rm max}=10^{13}{\rm M}_{\odot}$ the models produce ${\rm (S/N)}>3$ and $\chi_{\rm r}\leq 1$.

\section{Acknowledgments}
ESP and ODM thank Cl\'{a}udia V. Rodrigues, Ronaldo E. de Souza, Jos\'{e} C.N. de Araujo, and Jos\'{e} A. de Freitas Pacheco for helpful discussions. The authors would like to thank the referee for helpful comments that we feel considerably improved the paper. ESP was financially supported by the Brazilian Agency CAPES, and ODM is partially supported by CNPq (grant 305456/2006-7).

\label{lastpage}


\begin{thebibliography}{99}
\bibitem[\protect\citeauthoryear{Abel, Bryan \& Norman}{2002}]{abel} Abel T., Bryan G.L., Norman M.L., 2002, Sci, 295, 93

\bibitem[\protect\citeauthoryear{Allen}{1997}]{a1}Allen, B. 1997, in J.-A. Marck, \& J.-P. Lasota, eds, Relativistic Gravitation and Gravitational Radiation, Cambridge University Press,Princeton, NJ p. 373

\bibitem[\protect\citeauthoryear{Allen \& Romano}{1999}]{a2} Allen B., Romano J.D., 1999, Phys. Rev. D, 59, 2001

\bibitem[\protect\citeauthoryear{Bardeen et al.}{1986}]{b2} Bardeen J.M., Bond J.R., Kaiser N., Szalay A.S., 1986, ApJ, 304, 15

\bibitem[\protect\citeauthoryear{Belczynski, Kalogera \& Bulik}{2002}]{belcz} Belczynski K., Kalogera V., Bulik T., 2002, ApJ, 572, 407

\bibitem[\protect\citeauthoryear{Buonanno et al.}{2005}]{buo05} Buonanno A., Sigl G., Raffelt G.G., Janka H.T., M\"{u}ller E., 2005, Phys. Rev. D, 72, 084001

\bibitem[\protect\citeauthoryear{Carr}{1980}]{carr} Carr B.J., 1980, A\&A, 89, 6

\bibitem[\protect\citeauthoryear{Carrol, Press \& Turner}{1992}]{c2} Carrol S.M., Press W.H., Turner E.L, 1992, ARA\&A, 30, 499

\bibitem[\protect\citeauthoryear{Chiosi \& Maeder}{1986}]{chiosi} Chiosi C., Maeder A., 1986, ARA\&A, 24, 329

\bibitem[\protect\citeauthoryear{Christensen}{1992}]{crs} Christensen N., 1992, Phys. Rev. D, 46, 5250

\bibitem[\protect\citeauthoryear{Copi}{1997}]{c4} Copi C.J., 1997, Apj, 487, 704

\bibitem[\protect\citeauthoryear{Daigne et al.}{2006}]{d1} Daigne F., Olive K.A., Silk J., Stoehr F., Vangioni E., 2006, Apj, 647, 773

\bibitem[\protect\citeauthoryear{de Araujo, Miranda \& Aguiar}{2000}]{d3} de Araujo J.C.N., Miranda O.D., Aguiar O.D., 2000, Phys. Rev. D, 61, 12, 124015

\bibitem[\protect\citeauthoryear{de Araujo, Miranda \& Aguiar}{2002}]{d4} de Araujo J.C.N., Miranda O.D., Aguiar O.D., 2002, MNRAS, 330, 651

\bibitem[\protect\citeauthoryear{de Araujo, Miranda \& Aguiar}{2004}]{d5} de Araujo J.C.N., Miranda O.D., Aguiar O.D., 2004, MNRAS, 348, 1373

\bibitem[\protect\citeauthoryear{de Araujo \& Miranda}{2005}]{d2} de Araujo J.C.N., Miranda O.D., 2005, Phys. Rev. D, 71, 12, 12703

\bibitem[\protect\citeauthoryear{de Freitas Pacheco}{1997}]{pach1} de Freitas Pacheco J.A., 1997, Astrop. Physics, 8, 21 

\bibitem[\protect\citeauthoryear{Efstathiou, Bond \& White}{1992}]{e1} Efstathiou G., Bond J.R., White S.D.M., 1992, MNRAS, 258, 1p

\bibitem[\protect\citeauthoryear{Ellison et al.}{2000}]{ellison} Ellison S., Songaila A., Schaye J., Petinni M., 2000, AJ, 120, 1175

\bibitem[\protect\citeauthoryear{Ferrari, Matarrese \& Schneider}{1999}]{f2} Ferrari V., Matarrese S., Schneider R., 1999, MNRAS, 303, 247

\bibitem[\protect\citeauthoryear{Ferrari, Miniutti \& Pons}{2003}]{ferr03} Ferrari V., Miniutti G., Pons J.A., 2003, MNRAS, 342, 629

\bibitem[\protect\citeauthoryear{Flanagan}{1993}]{f5} Flanagan E.E., 1993, Phys. Rev. D, 48, 2389

\bibitem[\protect\citeauthoryear{Fryer, Woosley \& Heger}{2001}]{f3} Fryer C.L., Woosley S.E., Heger A., 2001, Apj, 550, 372

\bibitem[\protect\citeauthoryear{Giovannini}{2009}]{gio} Giovannini M., 2009, preprint (astro-ph:0901.3026)

\bibitem[\protect\citeauthoryear{Gunn \& Peterson}{1965}]{gunn} Gunn J.E., Peterson B.A., 1965, Apj, 142, 1633

\bibitem[\protect\citeauthoryear{Heger \& Woosley}{2002}]{h1} Heger A., Woosley S.E., 2002, Apj, 567, 532

\bibitem[\protect\citeauthoryear{Hopkins}{2004}]{h2} Hopkins A.M., 2004, Apj, 615, 209

\bibitem[\protect\citeauthoryear{Hopkins}{2007}]{h3} Hopkins A.M., 2007, Apj, 654, 1175

\bibitem[\protect\citeauthoryear{Hulse \& Taylor}{1974}]{hulse1} Hulse R.A., Taylor J.H., 1974, ApJ, 191, L59

\bibitem[\protect\citeauthoryear{Hulse \& Taylor}{1975a}]{hulse2} Hulse R.A., Taylor J.H., 1975, ApJ, 195, L51

\bibitem[\protect\citeauthoryear{Hulse \& Taylor}{1975b}]{hulse3} Hulse R.A., Taylor J.H., 1975, ApJ, 201, L55

\bibitem[\protect\citeauthoryear{Jenkis et al.}{2001}]{j1} Jenkins A., Frenk C.S., White S.D.M., Colberg J.M., Cole S., Evrard A. E., Couchman H., Yoshida N.,  2001, MNRAS, 321, 372

\bibitem[\protect\citeauthoryear{Kauffmann \& Haehnelt}{2000}]{kh00} Kauffmann G., Haehnelt M., 2000, MNRAS, 311, 576

\bibitem[\protect\citeauthoryear{Kroupa}{2007}]{kroupa} Kroupa P., 2007, preprint (astro-ph:0703124)

\bibitem[\protect\citeauthoryear{Luki\'{c} et al.}{2007}]{luk} Luki\'{c} Z., Heitmann K., Habib S., Bashinsky S., Ricker P. M., 2007, ApJ, 671, 1160

\bibitem[\protect\citeauthoryear{Maggiore}{2000}]{mag1} Maggiore M., 2000, Phys. Rep., 331, 283

\bibitem[\protect\citeauthoryear{Marassi, Schneider \& Ferrari}{2009}]{msf09} Marassi S., Schneider R., Ferrari V., 2009, MNRAS, 398, 293

\bibitem[\protect\citeauthoryear{Miranda, de Araujo \& Aguiar}{2004}]{m1} Miranda O.D., de Araujo J.C.N., Aguiar O.D., 2004, Class. Quantum Grav., 21, S557

\bibitem[\protect\citeauthoryear{Peacock}{1999}]{p1} Peacock J.A., 1999, Cosmological Physics, Cambridge University Press, Cambrige, 682

\bibitem[\protect\citeauthoryear{Press \& Schechter}{1974}]{p2} Press W.H., Schechter P., 1974, Apj, 193, 425

\bibitem[\protect\citeauthoryear{Regimbau \& de Freitas Pacheco}{2006}]{regimbau} Regimbau T., de Freitas Pacheco J.A., 2006, ApJ, 642, 455

\bibitem[\protect\citeauthoryear{Salpeter}{1955}]{s1} Salpeter E.E., 1955, Apj, 121, 161

\bibitem[\protect\citeauthoryear{Salvadori, Schneider \& Ferrara}{2007}]{sal} Salvadori S., Schneider R., Ferrara A., 2007, MNRAS, 381, 647

\bibitem[\protect\citeauthoryear{Sandick et al.}{2006}]{sand1} Sandick P., Olive K.A., Daigne F., Vangioni E., 2006, Phys. Rev. D, 73, 104024

\bibitem[\protect\citeauthoryear{Scalo}{1986}]{s6} Scalo J., Fund. Cosm. Phys., 1986, 11, 1

\bibitem[\protect\citeauthoryear{Schmidt}{1959}]{sch1} Schmidt M., 1959, Apj, 129, 243

\bibitem[\protect\citeauthoryear{Schmidt}{1963}]{sch2} Schmidt M., 1963, Apj, 137, 758

\bibitem[\protect\citeauthoryear{Sheth \& Tormen}{1999}]{st} Sheth R.K., Tormen G., 1999, MNRAS, 308, 119

\bibitem[\protect\citeauthoryear{Songaila \& Cowie}{1996}]{songaila} Songaila A., Cowie L.L., 1996, AJ, 112, 335

\bibitem[\protect\citeauthoryear{Springel \& Hernquist}{2003}]{sprher} Springel V., Hernquist L., 2003, MNRAS, 339, 312

\bibitem[\protect\citeauthoryear{Stark \& Piran}{1986}]{sp86} Stark R.F., Piran T., 1986, in R. Ruffini, ed., Proc. Fourth Marcel Grossmann Meeting on General Relativity. Elsevier Science, Amsterdam, p.327

\bibitem[\protect\citeauthoryear{Suwa et al.}{2007}]{suwa} Suwa Y., Takiwaki T., Kotake K., Sato K., 2007, ApJ, 665, L43

\bibitem[\protect\citeauthoryear{Suwa et al.}{2007}]{suwa2} Suwa Y., Takiwaki T., Kotake K., Sato K., 2007, PASJ, 59, 771

\bibitem[\protect\citeauthoryear{Thorne}{1987}]{t1} Thorne K. P.,  1987,  in S.W. Hawking and W. Israel, eds, Three Hundred Years of Gravitation, Cambridge University Press, Cambridge, p. 330

\bibitem[\protect\citeauthoryear{Venkatesan}{2000}]{venk} Venkatesan A.,
2000, ApJ, 537, 55

\bibitem[\protect\citeauthoryear{Wilkins, Trentham, \& Hopkins}{2008}]{wilkins} Wilkins S.M., Trentham N., Hopkins A.M., 2008, MNRAS, 385, 687

\bibitem[\protect\citeauthoryear{}{}{}]{}

\end{thebibliography}
\end{document}